\documentclass[onecolumn,12pt]{revtex4}
\usepackage{natbib} %<- to chyba musi byc przed bibunits
\usepackage{bibunits}
\usepackage{color}

\usepackage{etoolbox} %to musi byc dla apptocmd

\makeatletter
\newcommand*{\newbibstartnumber}[1]{%
	\apptocmd{\thebibliography}{%
		\global\c@NAT@ctr #1\relax
		\addtocounter{NAT@ctr}{-1}%
	}{}{}%
}
\makeatother

\defaultbibliographystyle{natureAstro}
\usepackage{epsfig}
\usepackage[colorlinks=true,citecolor=blue,urlcolor=blue,linkcolor=blue]{hyperref}
\usepackage{times}

\usepackage[normalem]{ulem}

\usepackage{placeins}
\begin{document}

%\preprint{}

\title{Searching for topological defect dark matter with optical atomic clocks}


\author{P. Wcis{\l}o}
\email[]{piotr.wcislo@fizyka.umk.pl}
%\thanks{}
\affiliation{Institute of Physics, Faculty of Physics, Astronomy and Informatics, Nicolaus Copernicus University, Grudzi\c{a}dzka 5, PL-87-100 Toru\'n, Poland}

\author{P. Morzy\'nski}
\affiliation{Institute of Physics, Faculty of Physics, Astronomy and Informatics, Nicolaus Copernicus University, Grudzi\c{a}dzka 5, PL-87-100 Toru\'n, Poland}

\author{M. Bober}
\affiliation{Institute of Physics, Faculty of Physics, Astronomy and Informatics, Nicolaus Copernicus University, Grudzi\c{a}dzka 5, PL-87-100 Toru\'n, Poland}

\author{A. Cygan}
\affiliation{Institute of Physics, Faculty of Physics, Astronomy and Informatics, Nicolaus Copernicus University, Grudzi\c{a}dzka 5, PL-87-100 Toru\'n, Poland}

\author{D. Lisak}
\affiliation{Institute of Physics, Faculty of Physics, Astronomy and Informatics, Nicolaus Copernicus University, Grudzi\c{a}dzka 5, PL-87-100 Toru\'n, Poland}

\author{R. Ciury{\l}o}
\affiliation{Institute of Physics, Faculty of Physics, Astronomy and Informatics, Nicolaus Copernicus University, Grudzi\c{a}dzka 5, PL-87-100 Toru\'n, Poland}

\author{M. Zawada}
\affiliation{Institute of Physics, Faculty of Physics, Astronomy and Informatics, Nicolaus Copernicus University, Grudzi\c{a}dzka 5, PL-87-100 Toru\'n, Poland}


\date{\today}

%\pacs{33.70.Jg}
%33.70.Jg	Line and band widths, shapes, and shifts
%34.20.Gj	Intermolecular and atom-molecule potentials and forces

%\keywords{}


\maketitle

%\section{Introduction}

%detection of the tau neutrino \cite{Kodama2001}
\begin{bibunit}

\textbf{The total mass density of the Universe appears to be dominated by dark matter. However, beyond its gravitational interactions at the galactic scale, little is known about its nature \cite{Derevianko2014,Budker2014,Budker2015,Stadnik2014,StadnikPRL2015-201301,Stadnik2015arxiv2,Roberts2015arxiv,Hees2016}. Extensions of the quantum electrodynamics Lagrangian with dark-matter coupling terms \cite{Essig2013} may result in changes to Standard Model parameters. Recently, it was proposed that a network of atomic clocks could be used to search for transient signals of a hypothetical dark matter \cite{Derevianko2014} in the form of stable topological defects \cite{Vilenkin1985}. The clocks become desynchronized when a dark-matter object sweeps through the network. This pioneering approach, which is applicable only for distant clocks, is limited by the quality of the fibre links \cite{Calonico2015}. Here, we present an alternative experimental approach that is applicable to both closely spaced and distant optical atomic clocks \cite{Rosenband2008,Chou2010,Hinkley2013,LeTargat2013,Bloom2014,Godun2014,Ludlow2015,Nicholson2015,Ushijima2015,Huntemann2016,Nemitz2016} and benefits from their individual susceptibilities to dark matter \cite{Stadnik2015,Stadnik2016PRA}, hence not requiring fibre links. We explore a new dimension of astrophysical observations by constraining the strength of atomic coupling to the hypothetical dark-matter cosmic objects. Our experimental constraint exceeds the previous limits \cite{Olive2008}; in fact, it not only reaches the ultimate level expected to be achievable with a constellation of GPS atomic clocks \cite{Derevianko2014} but also has a large potential for improvement.}

The main components of an optical atomic clock are a sample of cold, trapped atoms that are isolated from the environment and a laser locked to an ultra-stable optical cavity \cite{Ludlow2015}. Optimal for this purpose are atomic species that posses an ultra-narrow optical transition, called a clock transition. The exceptionally small spectral width of this transition together with a large value of the frequency of the optical radiation result in spectroscopic measurement with a high accuracy that has already reached $10^{-18}$ \cite{Nicholson2015,Ushijima2015}. In the ideal case of perfectly isolated atoms, the frequency of the clock transition $\omega_0$ is dictated by the values of fundamental physical constants. In typical applications, clock transitions serve as the most stable frequency references available. In our experiment, however, the different sensitivities of the clock transition and the optical cavity to variations in the fundamental constants \cite{Stadnik2015,Stadnik2016PRA} enable a search for non-gravitational signatures of topological defect dark matter (TDM).

Our proposed concept for a dark matter (DM) search technique is illustrated in Fig.~\ref{fig:rys1}. The set-up consists of two co-located optical atomic clocks. In each of them, the frequency of the laser (Laser 1 and Laser 2 in Fig.~\ref{fig:rys1}) is tightly locked to the ultra-stable optical cavity (Cavity 1 and Cavity 2 in Fig.~\ref{fig:rys1}). The beam, after passing through the frequency shifter (FS1 and FS2 in Fig.~\ref{fig:rys1}), probes the trapped atoms. The shifter correction is actively controlled to keep the frequency of the beam locked to the clock transition. Therefore, the changes to the frequency correction reflect the changes to the frequency of the clock transition with respect to the cavity. We define the clocks’ readouts as the frequency corrections at FS1 and FS2, and we denote them by $r_1(t)$ and $r_2(t)$, respectively. When the Earth traverses a TDM object, for a nonzero DM-SM coupling, the presence of the DM will perturb the values of certain Standard Model (SM) parameters. In particular, we may expect a transient variation in the electromagnetic fine-structure constant $\alpha$ that can be expressed as $\delta\alpha/\alpha=\phi^2/\Lambda_\alpha^2$, where $\phi$ is the DM field and $\Lambda_\alpha$ is the energy scale (which inversely parametrizes the strength of the DM-SM coupling) \cite{Derevianko2014,Stadnik2015arxiv,Derevianko2016}. This will shift the frequency of the electronic clock transition with respect to the cavity and hence will directly manifest in both readouts, $r_1(t)$ and $r_2(t)$. The theory of the response of an atomic clock transition, with respect to an optical cavity, to a variation in $\alpha$ was recently described by Stadnik and Flambaum \cite{Stadnik2015,Stadnik2016PRA}, and its application in our experiment is discussed in the Methods section.

In the ideal case, without experimental or environmental noise, a single optical atomic clock would be sufficient to detect a transient signal from a hypothetical DM object. Under real experimental conditions, however, the DM signature is expected to be hidden by noise; hence, a potential signal from a DM object cannot be distinguished from other effects. One possible solution is to simultaneously monitor two independent channels. If any observed common event is much larger than a possible common component estimated from all known physical phenomena, then it may be associated with an as-yet-unknown interaction. In the bottom part of Fig.~\ref{fig:rys1}, we show how the readouts of two co-located clocks could respond to a cascade of TDM objects passing through the clocks. We denote the common component of the readouts $r_1(t)$ and $r_2(t)$ by $s(t)$. A positive verification of the DM-SM coupling is impossible when the magnitude of the other common effects cannot be quantified. Nevertheless, a measurement of the common component $s(t)$ can still provide a constraint on the transient variation of $\alpha$ and on the magnitude of the DM-SM coupling. This common component can be extracted by cross-correlating the readouts $r_1(t)$ and $r_2(t)$. The constraint on the transient variation of the fine-structure constant can be expressed as
\begin{equation}
\frac{\delta\alpha}{\alpha}<\frac{1}{K_\alpha}\frac{\sqrt{A_0/\eta_{\tiny{\textrm{T}}}}}{\omega_0},
\label{Eq:ConstrainOnAlpha}
\end{equation}
where $A_0$ is the amplitude of the cross-correlation peak, $K_\alpha$ is a sensitivity coefficient (equal to one in the non-relativistic case \cite{Stadnik2015,Stadnik2016PRA}), and $\eta_{\tiny{\textrm{T}}}$ is the ratio of  the overall DM signal duration to the length  of the cross-correlated readouts, $t_2-t_1$ (see the Methods section for details). The inequality given in (\ref{Eq:ConstrainOnAlpha}) can be translated into a constraint on the energy scale in the hypothetical DM-SM interaction Lagrangian:
\begin{equation}
\Lambda_\alpha>d^{1/2}\sqrt{\sqrt{\frac{\eta_{\tiny{\textrm{T}}}}{A_0}}\rho_{\tiny{\textrm{TDM}}}\hbar c K_\alpha \mathcal{T}v \omega_0},
\label{Eq:ConstrainOnLambda}
\end{equation}
where $d$ is the size of the defects, $\rho_{\tiny{\textrm{TDM}}}$ is the mean DM energy density, $\mathcal{T}$ is the time between consecutive encounters with DM defects, and $v$ is the relative speed of the topological defects (see the Methods section for details).

In our measurement, we used a system of two nearly identical optical lattice clocks \cite{Bober2015,Morzynski2015} placed approximately ten meters apart (see the Methods section for details). Both clocks shared the same reference optical cavity, which was a dominant source of the common signal. We observed, however, that the majority of this signal was in the low-frequency range. Therefore, we applied a high-pass filter with a frequency cut-off at $0.027$~Hz to both readouts; see Fig.~\ref{fig:rys2}. This frequency cut-off means that the sensitivity of the measurement was decreased for topological defects of thicknesses exceeding the radius of the Earth, under the assumption of a relative topological defect speed of $v=300$~km/s \cite{Derevianko2014}. The cross-correlation of the two signals is depicted in Fig.~\ref{fig:rys2}. Its amplitude $A_0$ is equal to $0.73$~Hz$^2$. The width of the cross-correlation peak reflects the spectral characteristics of the common noise, and the oscillating shape is a result of the applied filter and the fact that the common noise is dominant. The length of each of the two recordings, $t_2-t_1$, is equal to $45700$~s. Under the arbitrary assumption that the thickness of a defect $d$ is comparable to the size of the Earth ($10000$~km) and that $\mathcal{T}=45700$~s, we may use inequality~(\ref{Eq:ConstrainOnAlpha}) to obtain an estimate of $\delta\alpha/\alpha<7\times 10^{-14}$ for the constraint on the transient variation in the fine-structure constant. Note that after the application of the high-pass filter, the amplitude of the cross-correlation peak is quite insensitive to the length of the cross-correlated signals for $t_2-t_1>1/(0.027\textrm{ Hz})$. For instance, if we cross-correlate only a $100$~s long portion of the entire readout duration, then, independent of which portion we choose, the amplitude is $A_0<1.9$~Hz$^2$. However, for this choice of $t_2-t_1$ ($=100$~s), the $\eta_{\tiny{\textrm{T}}}$  parameter is considerably larger ($\eta_{\tiny{\textrm{T}}}=0.33$), and hence, the constraint inferred from inequality~(\ref{Eq:ConstrainOnAlpha}) is much stronger:
\begin{equation}
\frac{\delta\alpha}{\alpha}<5\times 10^{-15}.
\label{Eq:ConstrainOnAlphaValue}
\end{equation}
Strong constraints on the time variation of $\alpha$ were reported in Refs.~\cite{Rosenband2008,Godun2014}. However, it should be noted that the physical meanings of those constraints and of that reported in this Letter are entirely different. The limits reported in Refs.~\cite{Rosenband2008,Godun2014} are barely sensitive to short transient effects.

We can also use our measurement to estimate a constraint on the energy scale $\Lambda_\alpha$ under the assumption of $\rho_{\tiny{\textrm{TDM}}}=0.3$ GeV cm$^{-3}$ \cite{Beringer2012}. Similarly to the considerations regarding $\alpha$, the results here also depend on the choice of $t_2-t_1$. The constraints on $\Lambda_\alpha$ for the entire readout duration, $t_2-t_1=45700$~s, and for a small portion thereof, $t_2-t_1=100$~s, are shown as functions of $d$ as the grey and green lines, respectively, in Fig.~\ref{fig:rys3}. The scaling of these constraints with the defect size is proportional to $d^{3/4}$.

The dashed blue and red lines in Fig.~\ref{fig:rys3} are the idealized constraints presented in a previous proposal \cite{Derevianko2014} for a trans-continental network of Sr optical lattice clocks and a GPS constellation, respectively. However, the sensitivity coefficient $K_\alpha$ was underestimated in Ref.~\cite{Derevianko2014} by a factor of $34$ (see the Methods section). The dotted blue and red lines in Fig.~\ref{fig:rys3} represent the corrected constraints. Thus, it can be seen that our experimental limit already reaches the constraint for the constellation of GPS clocks, even though Ref.~\cite{Derevianko2014} assumes more optimistic conditions, i.e., $\mathcal{T}=1$~yr. The green dashed line in Fig.~\ref{fig:rys3} depicts the possible constraint achievable using our method given the same ideal conditions considered in Ref.~\cite{Derevianko2014}, i.e., $\mathcal{T}=1$~yr and $\sigma/\omega_0=10^{-18}$ (for a $1$~s acquisition step). For this purpose, we assume $\eta=1/3$ for any $d$. The interpretations of these two approaches are, however, not exactly the same. Our approach provides a direct recipe for how to treat the experimental readouts, whereas that presented in the pioneering work of Derevianko and Pospelov \cite{Derevianko2014} is a general concept based on a phase-difference measurement. Furthermore, our constraint is valid for both distant and closely spaced clocks. This implies that even in the case of distant clocks, our method is not limited by the need for a phase-noise-compensated optical fibre link  of a length scale comparable to the size of the Earth \cite{Calonico2015} and that our limits presented in Fig.~\ref{fig:rys3} are therefore directly applicable.

The present experiment was performed using two co-located optical clocks. However, the same approach can be applied to distant clocks. For example, if two clocks were to be placed on opposite sides of the Earth and a series of TDM objects were then to traverse the Earth, moving from one clock to other (at  $v=300$~km/s), then the cross-correlation peak in Fig.~\ref{fig:rys2} would appear not at $\Delta t=0$~s but at $\Delta t\approx 42$~s. In a real experiment, the problem is more complex. For long measurements, the effects related to the translational and rotational motion of the Earth must be disentangled. To account for these effects, in one of the signals in the cross-correlation function, equation~(\ref{Eq:CrossCorrelation}), the time $t$ must be replaced with its corrected value
\begin{equation}
t\leftarrow t+f(\vec{r}(t),\vec{\theta}(t)),
\label{Eq:DistantClocks}
\end{equation}
where $f$ is a function of the position $\vec{r}$ and orientation $\vec{\theta}$ of the Earth with respect to a reference frame. For example, if we assume that the TDM objects are traversing the Earth along the equatorial plane, then the Earth's rotation can be disentangled from the signal by taking $f(t)=(2R_\oplus/v)\cos(2\pi t/T_{\tiny{\textrm{day}}})$, where $R_\oplus$ is the Earth's radius and $T_{\tiny{\textrm{day}}}$ is its period of rotation.

In our approach, the readouts of the distant clocks do not need to be correlated in real time. Just as in the standard radio-astronomical VLBI technique, they can be locally recorded (with time stamps that are accurate to the level of $1$~ms) and cross-correlated later. Any number of scenarios of the movement of TDM objects with respect to the Earth, and their associated $f$ functions, can be tested during data post-processing. For the correct function, the cross-correlation peak would be expected to occur at $\Delta t=0$. Otherwise, the common DM-SM interaction pulses would be mismatched and would destructively interfere in the cross-correlation response. Such a measurement with distant clocks could provide richer information about the motion of TDM objects. However, it would also require a correct guess regarding the form of the $f$ function. Therefore, for first attempts to detect the DM-SM coupling, the closely spaced arrangement seems more straightforward.

Although the origins of the common apparatus and environmental effects are different for the distant and co-located arrangements, the problem of estimating the magnitudes of these effects arises for both. Moreover, a common weakness of all such searches for hypothetical transient effects is the assumption that some minimal number of expected events will occur during the observation time. Therefore, it will be crucial to establish a long-term, worldwide observation program to continuously record the signals over years or even decades.

\end{bibunit}

\section*{Acknowledgements}

We are very grateful to Victor Flambaum and Yevgeny Stadnik for discussions and crucial remarks concerning the response of an atomic clock transition and an optical cavity to variations in the fine-structure constant, which helped us to properly evaluate our constraints. We also thank Wim Ubachs and Szymon Pustelny for inspiring discussions. The reported measurements were performed at the National Laboratory FAMO in Toru\'n, Poland, and were supported by a subsidy from the Ministry of Science and Higher Education. Support has also been received from the project EMPIR 15SIB03 OC18. This project has received funding from the EMPIR programme co-financed by the Participating States and from the European Union’s Horizon 2020 research and innovation programme. The individual contributors were partially supported by the National Science Centre of Poland through Projects Nos. 2015/19/D/ST2/02195, DEC-2013/09/N/ST4/00327, 2012/07/B/ST2/00235, DEC-2013/11/D/ST2/02663, 2015/17/B/ST2/02115, and 2014/15/D/ST2/05281. This research was partially supported by the TEAM Programme of the Foundation for Polish Science, co-financed by the EU European Regional Development Fund and by the COST Action CM1405 MOLIM. PW is supported by the Foundation for Polish Science's START Programme.

\section*{Author contributions}
P.W. developed the concept, performed the calculations and data analysis, and prepared the manuscript.  P.M. and M.B. performed the experiment. M.B., P.M., M.Z., D.L., A.C., and R.C. contributed to the development of the experimental set-up. P.W., R.C., M.Z., M.B. and P.M. contributed to the interpretation and discussion of the results. R.C., M.Z., D.L., P.M. and M.B. contributed to the preparation of the manuscript. M. Z. leads the experimental group.

\section*{Competing financial interests}
The authors declare no competing financial interests.

\begin{figure}[!h]
	\centering
	\includegraphics[width=0.8\columnwidth]{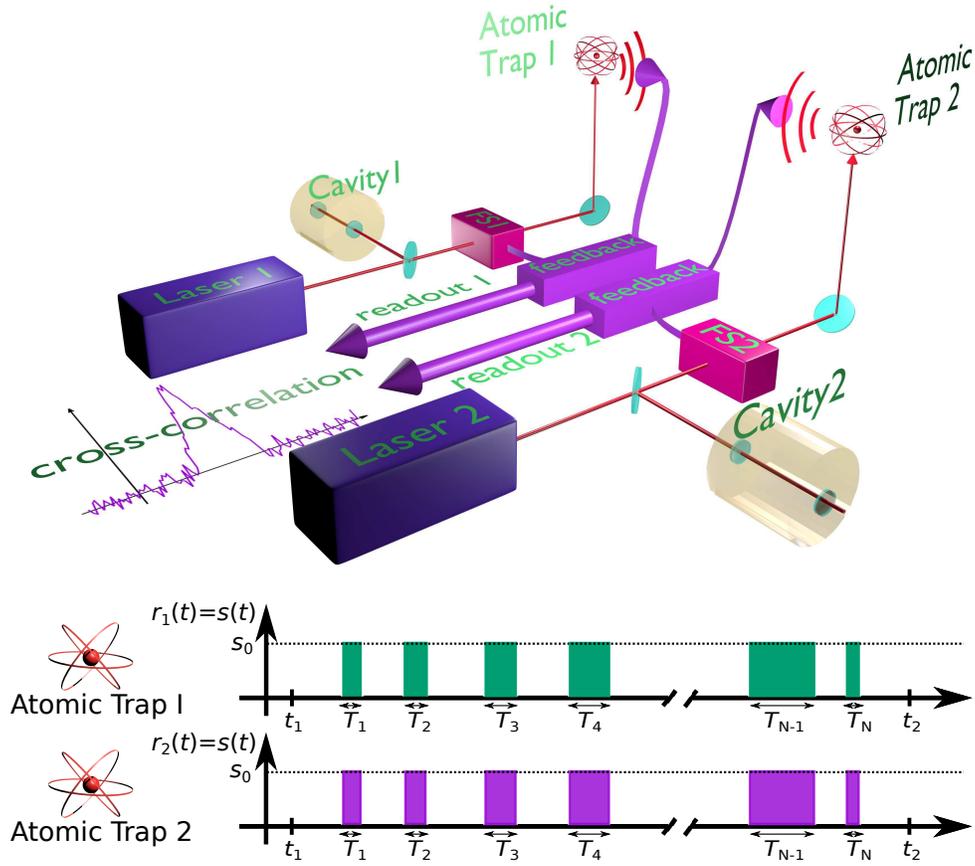} 
	\caption{\textbf{Set-up for a dark-matter (DM) search using two co-located optical atomic clocks.}  In each clock, the frequency of the probe laser (Laser 1 and Laser 2) is tightly locked to the ultra-stable optical cavity (Cavity 1 and Cavity 2). The frequency of the light is subsequently tuned by the frequency shifter (FS1 and FS2). The beam probes the clock transition in the trapped atoms (Atomic Trap 1 and Atomic Trap 2). The observed transition probability is used to generate feedback to actively control the shifter to keep the frequency of the probe beams locked to the clock transition. The changes to the frequency correction from each shifter (readout 1 and readout 2) reflect the changes to the frequency of the clock transition with respect to each cavity. \\ When the Earth traverses through a DM topological defect, the presence of the DM will perturb, given a non-zero DM-SM coupling, the values of certain Standard Model (SM) parameters. In particular, we may expect a transient variation in the fine-structure constant $\alpha$. This will shift the frequency of the electronic clock transition and hence will directly manifest as a common component $s(t)$ in both readouts, $r_1(t)$ and $r_2(t)$. In the simplest case, $s(t)$ can be modelled as a cascade of $N$ pulses of the same amplitude $s_0$ and durations of $T_1$, \dots, $T_N$.  The amplitude of the peak of the cross-correlation of $r_1(t)$ and $r_2(t)$ gives a constraint on the strength of the DM-SM coupling ($t_2-t_1$ is the length of the cross-correlated readouts).}
	\label{fig:rys1}
\end{figure}

\begin{figure}[!h]
	\centering
	\includegraphics[width=0.8\columnwidth]{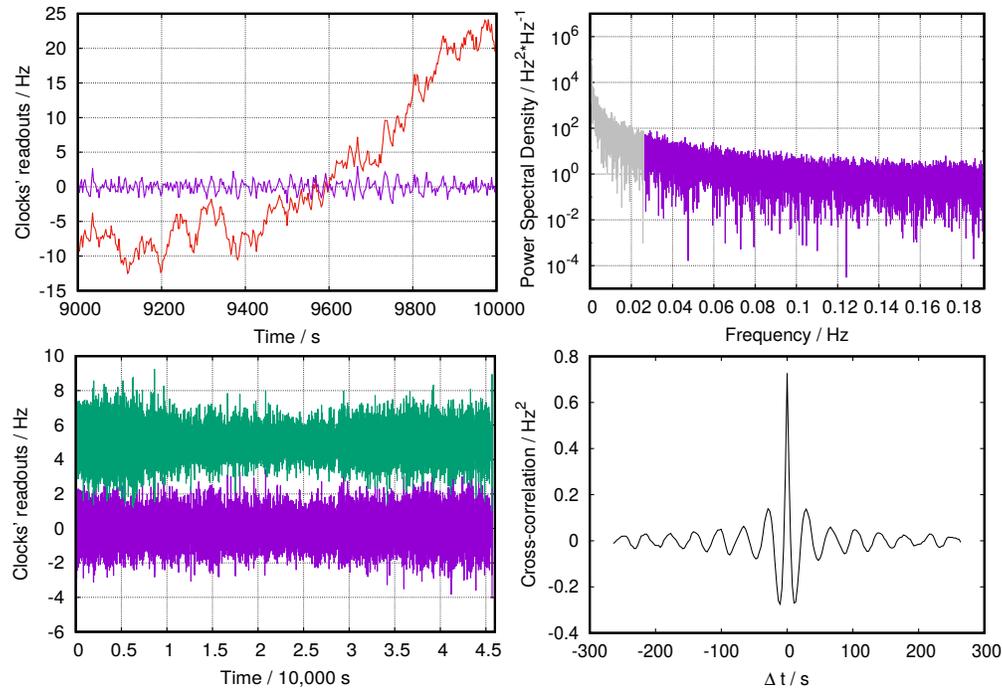} 
	\caption{\textbf{Extraction of the common signal from the clock's readouts.} The magnitude of the common component is retrieved by   cross-correlating the two readouts. First, a high-pass filter is applied to both readouts to eliminate the low-frequency common signal, which originates primarily from cavity instability. The top left panel shows a $1000$~s interval of the original (red) and filtered (purple) readouts from one of the clocks. The corresponding power spectrum density is presented in the top right panel. The dimmed area indicates the frequency range that is removed by the high-pass filter. The bottom left panel shows the entire readout for each clock after filtering (one of them is  artificially shifted for readability). The cross-correlation of the filtered readouts is shown in the bottom right panel. }
	\label{fig:rys2}
\end{figure}

\begin{figure}[!h]
	\centering
	\includegraphics[width=0.8\columnwidth]{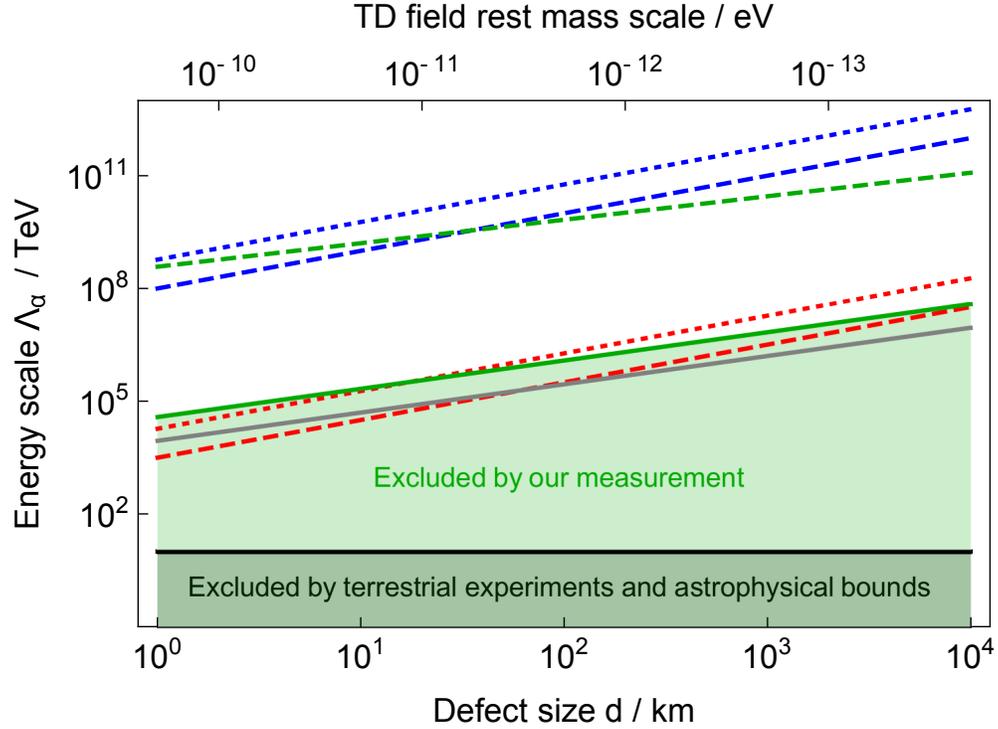} 
	\caption{\textbf{Constraint on the energy scale $\Lambda_\alpha$.} The grey and green lines represent the constraints we infer from the present measurement by cross-correlating the readouts in their entirety and by cross-correlating only a small portion of the readouts, respectively (see the text). The dashed blue and red lines, which are taken directly from Ref.~\cite{Derevianko2014}, represent limits on the experimental constraints that could be achieved with a trans-continental network of Sr optical lattice clocks and a GPS constellation, respectively. However, these limits are underestimated. The dotted blue and red lines represent the corrected constraints (see the Methods section for details). The dashed green line represents the limit achievable using our approach under the same conditions considered in Ref.~\cite{Derevianko2014}. }
	\label{fig:rys3}
\end{figure}

\begin{bibunit}
\newbibstartnumber{31}

\FloatBarrier

\section*{Methods}

\subsection*{Susceptibility of atoms and cavities to dark matter}

The expansion and cooling of the early Universe could have involved phase transitions related to hypothetical DM cosmic fields, which may have resulted in the formation of topological defects stabilized by a self-interaction potential. Such objects, if sufficiently light, would be of macroscopic size. We refer the reader to the review by Vilenkin \cite{Vilenkin1985}. Recently, several estimates concerning these possible objects have been presented by Derevianko and Pospelov; see the Supplementary Information to Ref.~\cite{Derevianko2014}. Extensions of the QED Lagrangian with DM-SM couplings (so-called portals \cite{Essig2013}) may result in variations in Standard Model parameters. For instance, the relative change in $\alpha$, for the case of a quadratic scalar portal, can be expressed as
\begin{equation}
\frac{\delta\alpha}{\alpha}=\frac{\phi_{\tiny{\textrm{inside}}}^2}{\Lambda_\alpha^2},
\label{Eq:AlphaVSEnergyScale}
\end{equation}
where $\phi_{\tiny{\textrm{inside}}}$ is the DM field inside a TDM object and $\Lambda_\alpha$ is the energy scale.

To consider how the electronic transition with respect to the optical cavity responds to the variation of $\alpha$, it is useful to write down the electronic Schr\"odinger equation in the Born-Oppenheimer approximation, in SI units, for $n$ electrons and $m$ nuclei:  
\begin{equation}
\left( -\frac{\hbar^2}{2 m_e}\sum\limits_{i=1}^{n}\nabla^2_{r_i}-\alpha\hbar c\sum\limits_{i,j=1}^{n,m}\frac{Z_j}{r_{ji}}+\frac{1}{2}\alpha\hbar c\sum\limits_{{\scriptsize \begin{array}{l}
	i,k=1\\
	i\neq k
	\end{array}}}^{n,n}\frac{1}{r_{ik}}\right) \psi=E\psi,
\label{Eq:schrodinger}
\end{equation}
where $m_e$ is the electron mass, $Z_j$ is the number of protons in the $j$-th nucleus, $r_{ji}=\left| \vec{R}_j -\vec{r}_i\right|$, and $r_{ik}=\left| \vec{r}_i -\vec{r}_k\right|$. The coordinates of the $j$-th nucleus and the $i$-th electron are $\vec{R}_j$ and $\vec{r}_i$, respectively. One may write equation~(\ref{Eq:schrodinger}) in dimensionless form as follows:
\begin{equation}
\left( -\frac{1}{2}\sum\limits_{i=1}^{n}\nabla^2_{x_i}-\sum\limits_{i,j=1}^{n,m}\frac{Z_j}{x_{ji}}+\frac{1}{2}\sum\limits_{{\scriptsize \begin{array}{l}
		i,k=1\\
		i\neq k
		\end{array}}}^{n,n}\frac{1}{x_{ik}}\right) \psi=\epsilon\psi,
\label{Eq:schrodinger2}
\end{equation}
where $\epsilon=E/E_h$ and $x_i=r_i/a_0$, $x_{ji}=r_{ji}/a_0$, and $x_{ik}=r_{ik}/a_0$, with $E_h=\alpha^2 m_e c^2$ and $a_0=\hbar/(m_e\alpha c)$. This simple transformation into the dimensionless form reveals a very important property, namely, that the values of the dimensionless energies $\epsilon$ and their dependence on the dimensionless positions of the nuclei $\vec{X}_j=\vec{R}_j/a_0$ do not depend on any fundamental constant. This means that, within this approximation, the energy of the system scales as $E\propto \alpha^2$ and its linear dimensions scale as $\propto \alpha^{-1}$ \cite{Stadnik2015,Stadnik2016PRA} for any system starting from simple atoms and molecules up to crystals and even amorphous solids. It turns out that this approximation works well for light elements (including those relevant to our experiment, i.e., Sr, Si, O and Ti) when relativistic effects can be neglected \cite{Angstmann2004,Flambaum2009}. In particular, the lengths of optical cavity spacers made either of single-crystal silicon \cite{Kessler2012} or completely amorphous Ultra-Low Expansion glass (as in our experiment) follow this simple $L \propto \alpha^{-1}$ scaling. Therefore, the frequency of the $N_{cav}$-th longitudinal cavity mode can be written as 
\begin{equation}
\omega_0^{cav}=N_{cav}c/(2L)\propto\alpha.
\label{Eq:freqCav}
\end{equation}
When relativistic effects are also considered, the dependence of the frequency of an electronic transition on the fine-structure constant $\alpha$ can be expressed, in SI units, as 
\begin{equation}
\omega_0^{Sr}\propto \alpha^{2+K_\alpha^{Sr}},
\label{Eq:freqSr}
\end{equation}
where $K_\alpha^{Sr}$ is a sensitivity factor \cite{Angstmann2004,Flambaum2009} that indicates how sensitive the transition is to $\alpha$ with respect to the non-relativistic case ($\omega_0\propto\alpha^2$). For light elements, for which relativistic effects are small, the sensitivity factor is close to zero. 

It follows from equations~(\ref{Eq:freqCav}) and (\ref{Eq:freqSr}) that the deviation of the frequency of the clock transition with respect to the cavity, $d\omega_0=d(\omega_0^{Sr}-\omega_0^{cav})$, due to the variation of $\alpha$ is
\begin{equation}
\frac{d\omega_0}{\omega_0}=K_\alpha\frac{d\alpha}{\alpha},
\label{Eq:FineStructureConst}
\end{equation}
where $K_{\alpha}=1+K_\alpha^X$ is the effective sensitivity coefficient, with $X$ representing the element used. For instance, in the case of our clocks, $K_\alpha^{Sr}=0.06$ \cite{Angstmann2004,Flambaum2009}; hence, the effective sensitivity is $K_\alpha=1.06$. For the much heavier element mercury, $K_\alpha=1.8$. In the non-relativistic case, $K_{\alpha}=1$, which reproduces the result presented by Stadnik and Flambaum \cite{Stadnik2015,Stadnik2016PRA}. In the above considerations based on SI units, we assume that the definition of the second is not perturbed by the presence of DM. 

It follows from equation~(\ref{Eq:freqSr}) that if we consider the arrangement proposed in Ref~\cite{Derevianko2014}, i.e., a comparison of clock transitions in two distant atomic samples, then the sensitivity coefficient would be $K_\alpha^{X}+2$, and not $K_\alpha^{X}$, as stated in Ref~\cite{Derevianko2014}. The constraints on $\Lambda_\alpha$ for optical and microwave atomic clocks that are given in Ref~\cite{Derevianko2014}, after this correction, are depicted as blue and red dotted lines, respectively, in Fig.~\ref{fig:rys3} of the main text.

\subsection*{Readout analysis}

The readouts from the two atomic clocks shown in Fig.~\ref{fig:rys1} (i.e., the frequency corrections from the shifters FS1 and FS2) may be written as $r_1(t)=n_1(t)+s(t)$ and $r_2(t)=n_2(t)+s(t)$, respectively, where $s(t)$ is the common component and $n_1(t)$ and $n_2(t)$ are the independent noise components in the two detectors. The common component can be extracted from the cross-correlation of the two readouts:
\begin{equation}
(r_1*r_2)(\Delta t)=\frac{1}{t_2-t_1}\int\limits_{t_1}^{t_2}r_1(t)r_2(t+\Delta t)dt,
\label{Eq:CrossCorrelation}
\end{equation}
where $t_1$ and $t_2$ are the limits of the time interval over which the signals were recorded. The cross-correlation, which is given by equation~(\ref{Eq:CrossCorrelation}), has a resonance-like shape centred at $\Delta t=0$ (see Fig.~\ref{fig:rys2}). The amplitude of this peak, $A_0=(r_1*r_2)(\Delta t=0~\textrm{s})$, reflects the magnitude of the common component $s(t)$. As a first approximation, let us assume that $s(t)$ consists of a series of $N_p$ square pulses with arbitrary durations $T_i$ and the same height $s_0$, which, in an ideal experiment, would correspond to DM objects of arbitrary dimensions and the same energy density $\rho_{\tiny{\textrm{inside}}}$ (see the bottom of Fig.~\ref{fig:rys1}). Then, the amplitude of the cross-correlation peak $A_0$ can be expressed as 
\begin{equation}
A_0=s_0^2\frac{T_{\tiny{\textrm{tot}}}}{t_2-t_1},
\label{Eq:CrossCorrelationAmplitude}
\end{equation}
where $T_{\tiny{\textrm{tot}}}=\sum\limits_{i=1}^{N_p}T_i$. If the two detectors do not suffer from any common instrumental noise or drifts (or they are negligible), then $s_0$ can be identified with the TDM signal $\delta \omega_0$. If the magnitude of the common apparatus effects cannot be quantified, then positive verification of the DM-SM coupling is impossible. Nevertheless, the amplitude of the cross-correlation peak still gives a constraint on its magnitude:  
\begin{equation}
\delta \omega_0 < \sqrt{A_0/\eta_{\tiny{\textrm{T}}}},
\label{Eq:Constrain}
\end{equation}
where $\eta_{\tiny{\textrm{T}}}=T_{\tiny{\textrm{tot}}}/(t_2-t_1)$.

It is also instructive to assume that the common apparatus effects are zero and ask how the detection limit of the system presented in Fig.~\ref{fig:rys1} depends on the intrinsic noise and stability of each of the two clocks, which we denote by $n_1(t)$ and $n_2(t)$. Both clocks are the same; hence, both noise components $n_1(t)$ and $n_2(t)$ are characterized by the same standard deviation $\sigma$. It follows from equation~(\ref{Eq:CrossCorrelation}) that the noise of the cross-correlation of the signals (its standard deviation) is $(\sigma/\sqrt{N})\sqrt{2\eta_{\tiny{\textrm{T}}} s_0^2+\sigma^2}$, where $N$ is the number of samples in the readout of each sensor. Under the assumption that either the signal is weak ($s_0\ll \sigma$) or the mean pulse duration is small ($\eta_{\tiny{\textrm{T}}}\ll 1$), only the leading term must be retained. If the common apparatus effects are negligible and no cross-correlation peak is observed, then a constraint on the strength of the DM-SM coupling can be determined from the noise level $\sigma$ of the clocks' readouts: 
\begin{equation}
\delta\omega_0 < \frac{\sigma}{\sqrt{2\eta_{\tiny{\textrm{T}}}\sqrt{N}}}.
\label{Eq:Constrain2}
\end{equation}
Inequality~{(\ref{Eq:Constrain2})} expresses a fundamental limitation of our approach. The important point that should emerge from this discussion is that the fundamental detection limit of our approach is limited not by the system instability, i.e., the minimum of the Allan variance, but only by the standard deviation of the clock noise at short time scale. Indeed, the goal of this experiment is not an ultra-accurate determination of the frequency of the clock transition in the two sensors but merely the possibly precise tracking of their common changes. Therefore, the readouts can be arbitrarily long, even exceeding the length of the period indicated by the minimum of the Allan variance, because clock drift is no longer an issue.

Inequalities~(\ref{Eq:Constrain}) and (\ref{Eq:Constrain2}) provide a complete recipe for interpreting the outcome of the experiment depicted in Fig.~\ref{fig:rys1}. If a peak is observed in the cross-correlation function (at $\Delta t=0$), then its amplitude determines a constraint on the effect of the DM according to inequality~(\ref{Eq:Constrain}). Otherwise, if the cross-correlation is flat, the constraint is determined by the standard deviation of a single measurement $\sigma$ and by the number of samples $N$, as shown in inequality~(\ref{Eq:Constrain2}). When the influence of the common apparatus effects is estimated to be negligible, a positive verification of the hypothetical DM-SM interaction is possible, and the strength of the TDM signal is $\delta\omega_0=\sqrt{A_0/\eta_{\tiny{\textrm{T}}}}$. The same approach may be implemented in other types of experiments looking for transient effects of dark matter, such as those based on synchronized optical magnetometers \cite{Pustelny2013,Pospelov2013}.

The constraint on the transient variation in $\alpha$ that is given by inequality~(\ref{Eq:ConstrainOnAlpha}) in the main text can be derived directly from equation~(\ref{Eq:FineStructureConst}) and inequality~(\ref{Eq:Constrain}). The constraint on the energy scale $\Lambda_\alpha$ of the hypothetical DM-SM interaction Lagrangian that is given by inequality~(\ref{Eq:ConstrainOnLambda}) of the main text can be derived by using equation~(\ref{Eq:AlphaVSEnergyScale}) and (following Derevianko and Pospelov \cite{Derevianko2014}) estimating the magnitude of the field inside a TDM object as  $\phi_{\tiny{\textrm{inside}}}^2=\mathcal{T}v\rho_{\tiny{\textrm{TDM}}}d \hbar c$.

\subsection*{Experimental set-up}

The experimental set-up consists of two optical atomic clocks \cite{Ludlow2015} with neutral $^{88}$Sr atoms trapped in optical lattices \cite{Takamoto2005}. Each clock cycle is divided into preparation and interrogation phases. In the preparation phase, using the $^1$S$_0$-$^1$P$_1$ transition at $461$~nm, hot strontium atoms from the oven are first cooled to $2$~mK using a Zeeman slower \cite{Phillips1982} and a magneto-optical trap \cite{Raab1987}. Next, by using the $^1$S$_0$-$^3$P$_1$ $7.5$~kHz intercombination transition, we further cool the atoms to a few $\mu$K. Subsequently, a sample of a few thousand atoms is loaded into a one-dimensional optical lattice operating at $813$~nm. In the interrogation phase, the strontium atoms are probed at the $^1$S$_0$-$^3$P$_0$ so-called clock transition at $698$~nm ($429$~THz) with the ultra-narrow clock laser stabilized to the optical cavity. This probing allows the excitation probability to be measured as a function of the probe light frequency and allows the feedback correction to this laser frequency to be calculated. This process heats atoms and removes them from the lattice.

In our set-up \cite{Bober2015, Morzynski2015}, the two optical atomic clocks share the same optical cavity. The laser that is locked to it is split into two beams, whose frequencies are controlled separately by acousto-optic frequency shifters. The optical cavity consists of a $100$~mm Ultra-Low Expansion (ULE) glass spacer and fused silica mirrors. The relative instability of the clock laser, measured using the strontium atoms, is lower than $2\times 10^{-14}$ on time scales of $2-10^4$~s.

The clock laser frequency is digitally stabilized to the atomic transition by feedback corrections applied to the frequency shifters. The frequency corrections are derived from the excitation probabilities recorded on both sides of the clock transition line using the algorithm described in, e.g., \cite{Morzynski2013}. One servo-loop cycle, i.e., a single readout, requires two cycles of the clock.

The presented results were measured with synchronized cycles (of $1.3$~s) of the two clocks, which means that the probing of the clock transition (with an interrogation time of $40$~ms) was performed in parallel in the two systems and corrections were applied simultaneously. The readouts from the two clocks were recorded independently by two different computers with their internal clocks synchronized to the UTC by a stratum 1 NTP server.

The interpretation of the clocks' response to TDM events is straightforward when their duration is longer than one servo-loop cycle ($2.6$~s in our case). However, in the limit of one servo-loop cycle, the sensitivity of the clocks to TDM objects could drop by a factor of $2$ for typical servo-loop settings. Nevertheless, if we assume that during the observation time, a sufficiently large number of TDM events occurs (randomly distributed in time), then our approach is also capable of detecting events shorter than one servo-loop cycle. In this case, on the one hand, the probability that such a TDM event will overlap with the interrogation period is smaller than one. On the other hand, the contribution of any single such TDM event to the height of the cross-correlation peak will be overestimated. These two effects cancel out, causing the sensitivity to be the same as that for longer events. For events that partially overlap the interrogation period, we assume that the clocks' response is proportional to the overlap fraction; however, the signal may vary slightly depending on the type of spectroscopy used (in our case, Rabi spectroscopy).

\subsection*{Distant clocks}

In the proposal presented in Ref.~\cite{Derevianko2014}, the clock transitions in the two distant atomic samples are directly compared via an optical link. Calonico et al. \cite{Calonico2015} have demonstrated that phase-noise-compensated optical fibre links of length scale comparable to the size of the Earth, as would be required for this purpose, constitute a limiting factor for such measurements. Our approach does not suffer from this source of noises because it does not require such an optical link to operate. Indeed, as shown in Fig.~\ref{fig:rys1} of the main text, there is no optical connection between the two channels. In our approach, each channel possesses two optical references, i.e., the atomic clock transition and the optical cavity. Therefore, the comparison of these two frequencies can be performed locally and separately for each channel. The readouts from the two channels can then be analysed digitally during data post-processing. In our approach, the clocks' cycles of operation must be synchronized at the millisecond level, which is simple to achieve using GPS or a standard internet connection. Here, the discussion is restricted to a set-up consisting of two optical atomic clocks; however, these remarks also apply to a system composed of a larger number of detectors.\\

All data supporting the findings of this study are available from the authors upon reasonable request.
%\putbib	

\end{bibunit}

%\bibliographystyle{apsrev4-1}
%\bibliography{DM2}

\end{document}